\documentclass{iopart}
\usepackage[a4paper, top=3cm, bottom=3cm, left=2cm, right=2cm]{geometry}
\usepackage{iopams}
\expandafter\let\csname equation*\endcsname\relax
\expandafter\let\csname endequation*\endcsname\relax
\usepackage{amsmath}
\usepackage{amssymb}
\usepackage{textcomp}
\usepackage{graphicx}
\usepackage{multicol}
\usepackage{bbm}
\newcommand{\ens}[0]{\ensuremath}
\newcommand{\iE}[0]{\ens{\mathrm{i}}}
\newcommand{\Eins}[0]{\ens{\mathbbm{1}}} %use mathds instead?

\usepackage[backend=bibtex,style=numeric,sorting=none,firstinits,url=false,maxnames=99]{biblatex}
\bibliography{Draft6_Single.bib}
\begin{document}

\title{Holographic quantum imaging: reconstructing spatial properties via two-particle interference}
\author{Nils Trautmann}
\address{Institut f{\"u}r Angewandte Physik, Technische Universi{\"a}t Darmstadt, D-64289 Darmstadt, Germany\\\emph{and}}
\address{School of Physics and Astronomy, SUPA, University of Glasgow, 
Glasgow G12 8QQ, UK}
\author{Gergely Ferenczi}
\address{School of Physics and Astronomy, SUPA, University of Glasgow, 
Glasgow G12 8QQ, UK}
\author{Sarah Croke}
\address{School of Physics and Astronomy, SUPA, University of Glasgow, 
Glasgow G12 8QQ, UK}
\author{Stephen M. Barnett}
\address{School of Physics and Astronomy, SUPA, University of Glasgow, 
Glasgow G12 8QQ, UK}
\begin{abstract}
Two particle interference phenomena, such as the Hong-Ou-Mandel effect, are a direct manifestation of the nature of the symmetry properties of indistinguishable particles as described by quantum mechanics. The Hong-Ou-Mandel effect has recently been applied as a tool for pure state tomography of a single photon. In this article, we generalise the method to extract additional information for a pure state and extend this to the full tomography of mixed states as well. The formalism is kept general enough to apply to both boson and fermion based interferometry. Our theoretical discussion is accompanied by two proposals of interferometric setups that allow the measurement of a tomographically complete set of observables for single photon quantum states. 
\end{abstract}

\date{\today}

\maketitle

\section{Introduction\label{sec:Introduction}}
%\begin{multicols}{2}
\noindent The concept of indistinguishable particles lies at the heart of quantum mechanics and quantum statistics. Two particle interference phenomena such as the Hong-Ou-Mandel effect (HOM) \cite{HOM} are a direct manifestation of the quantum mechanical description of indistinguishable particles. As such two particle interference effects provide fundamental tests of the foundations of quantum mechanics. In recent years the HOM effect has become a very useful tool throughout quantum optics. It has been used for generating entangled states \cite{moehring2007entanglement}, performing Bell state measurements \cite{michler1996interferometric} and testing the preparation of indistinguishable photon pairs \cite{santori2002indistinguishable} amongst other things. These effects emerge for example when two identical particles are incident on distinct input ports of a balanced beam splitter. Two identical bosonic particles such as photons will always leave in the same output port, this is known as HOM effect \cite{HOM}, while two identical fermions in the same scenario always leave in distinct output ports \cite{Yamamoto}.\\
HOM interference for identical bosons and for identical fermions contrasts the behaviour of states symmetric and antisymmetric (respectively) under the exchange of entrance or exit port mode numbers \cite{Loudon}. This effect can be used to interrogate the exchange symmetry of the state of a second degree of freedom of the particle pair \cite{MultimodeHOM}. This has been demonstrated for polarization, \cite{T76} and forms the basis of partial Bell state analysis, and for the orbital angular momenta of photons \cite{OAMHOM} where it has been used to sort the entangled state resulting from down conversion according to the parity of the angular momenta.\\
A recent experiment by Chrapkiewicz et al. \cite{HOMTomography} applies HOM interference as a quantum imaging technique for a pure single photon state. The unknown photon and a reference photon (known entirely and under experimental control) are interfered on a beam splitter and imaged onto a detector array. In this case, the interrogated degree of freedom is the transverse spatial profile of the photons. For a given pair of transverse spatial modes, the difference in count rates between the photons being detected in the same port and the photons being detected in distinct output ports provides information about the relative phase of the amplitudes related by the exchange of transverse spatial modes. As the state of the reference photon is known entirely this translates into obtaining information about the relative phase between two spatial modes of the unknown photon. \\
In this article, we provide an in-depth analysis of this imaging technique applied in \cite{HOMTomography}. We generalise this by showing that with carefully engineered loss or the inclusion of additional degrees of freedom, such as polarization, one can gain access to observables whose measurements provide tomographically complete information about mixed single particle quantum states. We work in the second quantised formalism keeping the analysis open enough to apply to both bosonic and fermionic particles. Our theoretical discussion of the protocol is accompanied by two proposals of interferometric setups for performing the tomography for mixed single photon states.\\
This paper is organized as follows. In Sec. \ref{sec:Model} the model of an interferometric setup for two particle interference is presented. The technique for performing the state tomography for pure single particle states is discussed in Sec. \ref{sec:State_tomography_pure_state}. We analyse the conditions for generalizing the scheme to provide a protocol for performing the state tomography for mixed states in Sec. \ref{sec:State_tomography_mixed_states}. In Sec. \ref{sec:Photonic_Implementation} we show that these conditions can be satisfied by proposing two interferometric setups for performing the state tomography for mixed single photon states.
\section{Physical model\label{sec:Model}}
\begin{figure*}
\begin{centering}
\includegraphics[width=6cm]{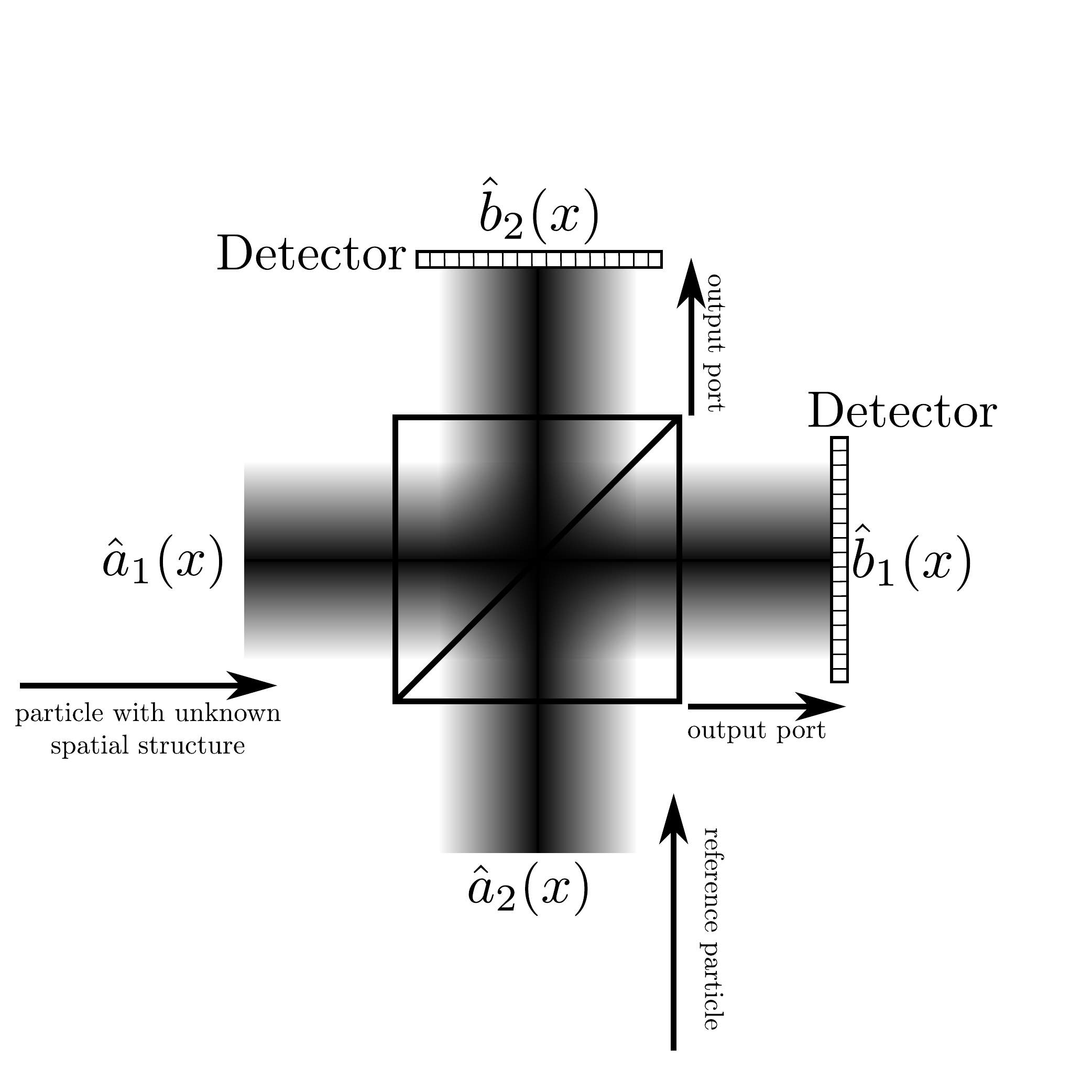} 
\par\end{centering}
\caption{Basic setup for two particle interferometer. We assume that the reference particle (known entirely and under experimental control) is entering the beam splitter at input port $2$ and the particle with the unknown transverse spatial profile is entering the beam splitter at input port $1$. The output ports are imaged onto two detector arrays and the resulting coincidence counts on the detectors are measured. The annihilation operators at \emph{input} port $\alpha$ are denoted $\hat{a}_{\alpha}(x)$ and the annihilation operators at \emph{output} port $\alpha$ are denoted $\hat{b}_{\alpha}(x)$, with $x$ referring to the transverse position of the particle. \label{fig:Setup}}
\end{figure*}
We are considering two degrees of freedom. In the schematic model of the experiment the first degree of freedom consists of two spatial paths that interfere in a suitable imaging apparatus. The second degree of freedom, the one to be imaged, is the transverse spatial profile of the particles. Note that the underlying principle of the imaging technique discussed herein does not depend on what the two degrees of freedom are physically. While we keep the second degree of freedom as the transverse spatial profile throughout this text, the role of the first degree of freedom  is performed by polarization modes in the implementation presented in section \ref{subsec:Loss} and by the four-mode degree of freedom formed by the port and polarization modes in the implementation presented in section \ref{subsec:Polarization}.\\
The annihilation operator at \emph{input} port $\alpha$ is denoted $\hat{a}_{\alpha}(x)$ and the annihilation operator at \emph{output} port $\alpha$ is denoted $\hat{b}_{\alpha}(x)$, with $x$ referring the transverse position of the particle. We restrict our analysis to one transverse dimension for notational simplicity, but introducing both transverse dimensions represents no inherent difficulty. The corresponding creation operators are used to describe quantum states in port $1$ or $2$ respectively as 
\begin{equation}
\vert x\rangle_{1/2}=\hat{a}_{1/2}^{\dag}(x)\vert0\rangle_{1/2}
\end{equation}
where $\vert 0 \rangle_{1/2}$ denotes the vacuum in port $1$ or $2$. We can describe the imaging system using the relation 
\begin{equation}
\left(\begin{array}{c}
\hat{b}_{1}(x)\\
\hat{b}_{2}(x)
\end{array}\right)=\underbrace{\left(\begin{array}{cc}
U_{11} & U_{12}\\
U_{21} & U_{22}
\end{array}\right)}_{=U}\left(\begin{array}{c}
\hat{a}_{1}(x)\\
\hat{a}_{2}(x)
\end{array}\right)\,,\label{eq:retrodiction-1}
\end{equation}
where the matrix $U$ characterizes the specific properties of the imaging system. In the absence of particle loss $U$ is a unitary matrix, i.e. $UU^{\dagger}=\Eins$. Losses can be described by introducing additional degrees of freedom to produce an effective non-unitary matrix \cite{LossyBS}. We shall return to this in section \ref{sec:Photonic_Implementation}. The input state consists of a particle prepared in an unknown state in port 1 and a reference particle in port 2 characterized by normalised transverse  spatial profiles $\psi_{u}(x)$ and $\psi_{r}(x)$ respectively 
%\begin{equation}
%\begin{split}\vert\psi_{\text{in}}\rangle_{1,2} & =\vert\psi_{u}\rangle_{1}\otimes\vert\psi_{r}\rangle_{2}\\
% & =\int_{\mathbb{R}}\int_{\mathbb{R}}dxdy\psi_{u}\left(x\right)\psi_{r}\left(y\right)\hat{a}_{1}^{\dagger}(x)\hat{a}_{2}^{\dagger}(y)\vert0\rangle_{1,2}.
%\end{split}
%\label{eq:Input}
%\end{equation}
\begin{equation}
\begin{split}
\vert\psi_{\text{in}}\rangle_{1,2} & =\vert\psi_{u}\rangle_{1}\otimes\vert\psi_{r}\rangle_{2}\\
\end{split}
\label{eq:Input}
\end{equation}
with
\begin{align}
\vert\psi_{u}\rangle_{1} & =\int_{\mathbb{R}}dx\psi_{u}\left(x\right)\vert x\rangle_{1}\\
\vert\psi_{r}\rangle_{2} & =\int_{\mathbb{R}}dx\psi_{r}\left(x\right)\vert x\rangle_{2}.
\end{align}
Information is gathered about this input state by joint spatially resolved detections at the output ports. The quantity of interest is 
\begin{equation}
p_{\alpha,\beta}(x,y)=\vert_{1,2}\langle\psi_{x,y}^{\alpha,\beta}\vert\psi_{\mathrm{in}}\rangle_{1,2}\vert^{2}\label{eq:Probability}
\end{equation}
where $p_{\alpha,\beta}(x,y)dxdy$ is the joint probability of detecting a particle in port $\alpha$ between $x$ and $x+dx$ and of detecting a particle in port $\beta$ between $y$ and $y+dy$. The state $\vert\psi_{x,y}^{\alpha,\beta}\rangle$ is the two particle state corresponding to this detection event 
\begin{equation}
\vert\psi_{x,y}^{\alpha,\beta}\rangle_{1,2}=\hat{b}_{\alpha}^{\dag}(x)\hat{b}_{\beta}^{\dag}(y)\vert0\rangle_{1,2}\label{eq:DetectionState12}
\end{equation}
and hence shall be referred to as the detection state. The probabilities $p_{\alpha,\beta}(x,y)$ need to be treated with some care as they include, implicitly, the physically \emph{indistinguishable possibilities} $p_{\beta,\alpha}(y,x)$. The sum of the probabilities for \emph{exclusive} events must be unity and so it is necessary to count these two contributions only once. The easiest way to do this is to require that
\begin{equation}
\int_{-\infty}^\infty dx\int_x^\infty dy\sum_{\alpha,\beta\in\{1,2\}}p_{\alpha,\beta}(x,y) = 1.
\end{equation}
Careful calculation (as shown in section \ref{sec:State_tomography_pure_state}) confirms that this is indeed the case for the probabilities \eqref{eq:Probability}.\\
At this point the problem is to reconstruct a two-particle state based on the detection probabilities \eqref{eq:Probability}. However, we can consider the reference particle as part of the detection mechanism so that we are left with the problem of performing tomography on a single particle state $\vert\psi_{u}\rangle_{1}$. Formally this can be achieved by writing the detection state in terms of the input modes, using equation \eqref{eq:retrodiction-1}, and taking the overlap of the state across the two input ports so obtained with the state of the reference particle in input port 2 
\begin{align}
_{1}\langle\psi_{x,y}^{\alpha,\beta}\vert  =&{}_{1,2}\langle\psi_{x,y}^{\alpha,\beta}\vert\psi_{r}\rangle_{2}\\
  =&{}_{1,2}\langle0\vert\sum_{\mu,\nu\in\{1,2\}}U_{\beta\mu}U_{\alpha\nu}\hat{a}_{\mu}(y)\hat{a}_{\nu}(x)\int_{\mathbb{R}}dz\psi_{r}(z)\hat{a}^\dag_{2}(z)\vert0\rangle_{2}\\
  =&{}_{1}\langle0\vert\left(U_{\beta1}U_{\alpha2}\psi_{r}(x)\hat{a}_{1}(y)\pm U_{\beta2}U_{\alpha1}\psi_{r}(y)\hat{a}_{1}(x)\right)\label{eq:DetectionState1Bra}
\end{align}
thereby turning all components of the detection state in port 2 into complex numbers and resulting in a state purely in port 1. Thus the problem is reduced to probing the unknown state in port 1 
\begin{equation}
p_{\alpha,\beta}(x,y)=\vert_{1}\langle\psi_{x,y}^{\alpha,\beta}\vert\psi_{u}\rangle_{1}\vert^{2}\label{eq:ProbabilityReduced}
\end{equation}
with the detection state, now also in only port 1, taking the form
\begin{equation}
\vert\psi_{x,y}^{\alpha,\beta}\rangle_{1}=U_{\beta1}^{*}U_{\alpha2}^{*}\psi_{r}^{*}(x)\vert y\rangle_{1}\pm U_{\beta2}^{*}U_{\alpha1}^{*}\psi_{r}^{*}(y)\vert x\rangle_{1}.\label{eq:DetectionState1}
\end{equation}
In arriving at result \eqref{eq:DetectionState1Bra} the (anti)commutation relations 
\begin{equation}
\hat{a}_{\alpha}(x)\hat{a}_{\beta}^{\dag}(y)\mp\hat{a}_{\beta}^{\dag}(y)\hat{a}_{\alpha}(x)=\delta_{\alpha\beta}\delta(x-y)\label{eq:(Anti)Commutator}
\end{equation}
have been used with the upper sign describing bosons and the lower sign describing fermions. \\

\section{State tomography for pure states\label{sec:State_tomography_pure_state}}
The resulting probability densities \eqref{eq:ProbabilityReduced}, to be used throughout the following, are
\begin{align}
p_{\alpha,\beta}\left(x,y\right)  = & \left|U_{\beta1}U_{\alpha2}\psi_{r}(x)\psi_{u}(y)\pm U_{\beta2}U_{\alpha1}\psi_{r}(y)\psi_{u}(x)\right|^{2}\nonumber \\
  = & \left|U_{\beta1}U_{\alpha2}\right|^{2}\left|\psi_{r}(x)\right|^{2}\left|\psi_{u}(y)\right|^{2}\nonumber \\
   & +\left|U_{\beta2}U_{\alpha1}\right|^{2}\left|\psi_{r}(y)\right|^{2}\left|\psi_{u}(x)\right|^{2}\nonumber \\
   & \pm2\text{Re}\left[U_{\beta1}U_{\alpha2}U_{\beta2}^{*}U_{\alpha1}^{*}\psi_{r}(x)\psi_{u}^{*}(x)\psi_{u}(y)\psi_{r}^{*}(y)\right]\nonumber \\
\label{eq:ProbabilityExplicit}
\end{align}
%\begin{small}
%\begin{eqnarray}
%p_{\alpha,\beta}\left(x,y\right) & = & \left|U_{\beta1}U_{\alpha2}\psi_{r}(x)\psi_{u}(y)\pm U_{\beta2}U_{\alpha1}\psi_{r}(y)\psi_{u}(x)\right|^{2}\nonumber \\
% & = & \left|U_{\beta1}U_{\alpha2}\right|^{2}\left|\psi_{r}(x)\right|^{2}\left|\psi_{u}(y)\right|^{2}\nonumber \\
% &  & +\left|U_{\beta2}U_{\alpha1}\right|^{2}\left|\psi_{r}(y)\right|^{2}\left|\psi_{u}(x)\right|^{2}\nonumber \\
% &  & \pm2\text{Re}\left[U_{\beta1}U_{\alpha2}U_{\beta2}^{*}U_{\alpha1}^{*}\psi_{r}(x)\psi_{u}^{*}(x)\psi_{u}(y)\psi_{r}^{*}(y)\right]\nonumber \\
%\label{eq:ProbabilityExplicit}
%\end{eqnarray}
%\end{small}
Note that the probability density (prior to taking the modulus squared) is composed of a linear combination of the amplitudes of the position exchanged alternatives of the particle pair each of which is multiplied by the appropriate transition amplitudes. For a balanced beam splitter, described by \cite{schleich2011quantum}
\begin{equation}
U  =  \frac{1}{\sqrt{2}}\left(\begin{array}{cc}
1 & 1\\
1 & -1
\end{array}\right),\label{eq:Unitary_beam_splitter}
\end{equation}
the transition amplitudes $(U_{\alpha\gamma}U_{\beta\delta})$ in all cases have an absolute value of $1/2$ and the probabilities \eqref{eq:ProbabilityExplicit}
take on the rather simple form 
%\begin{align}
%p_{1,1}(x,y)=p_{2,2}(x,y) & =\frac{1}{4}\vert\psi_{r}(x)\psi_{u}(y)\pm\psi_{r}(y)\psi_{u}(x)\vert^{2}\label{eq:ProbabilitySpecialCase1}\\
%p_{1,2}(x,y)=p_{2,1}(x,y) & =\frac{1}{4}\vert\psi_{r}(x)\psi_{u}(y)\mp\psi_{r}(y)\psi_{u}(x)\vert^{2}.\label{eq:ProbabilitySpecialCase2}
%\end{align}
\begin{equation}
p_{\alpha,\beta}(x,y) = \frac{1}{4}\vert\psi_{r}(x)\psi_{u}(y)\pm(-1)^{\alpha-\beta}\psi_{r}(y)\psi_{u}(x)\vert^{2}\label{eq:ProbabilitySpecialCase}
\end{equation}
When \eqref{eq:ProbabilitySpecialCase} is expanded and summed over the port mode labels the cross terms cancel as there are exactly two sets of cross terms where $\alpha$ and $\beta$ are equal and exactly two sets where $\alpha$ and $\beta$ differ by 1. Hence 
\begin{equation}
\begin{split}
&\int_{-\infty}^\infty dx\int_x^\infty dy\sum_{\alpha,\beta\in{1,2}}p_{\alpha,\beta}(x,y) \\
&=\int_{-\infty}^\infty dx\int_x^\infty dy\left(\vert\psi_r(x)\vert^2\vert\psi_u(y)\vert^2 + \vert\psi_r(y)\vert^2\vert\psi_u(x)\vert^2 \right) \\
&=\int_{-\infty}^\infty dx\vert\psi_r(x)\vert^2\int_{-\infty}^\infty dy\vert\psi_u(y)\vert^2 = 1
\end{split}
\end{equation}
where we have used the fact that the two input particles are each normalised.\\
We see from \eqref{eq:ProbabilitySpecialCase} that, for bosons, the probabilities of ending up in the same port are determined by the symmetric combination of the position exchanged amplitudes and the probabilities for ending up in different ports are determined by the antisymmetric combination of the position exchanged amplitudes. The converse is true for fermions. If we are dealing with a reference particle with a flat profile in the relevant region of space, i.e. 
\begin{equation}
\psi_{r}(x)=c\,.
\end{equation}
then the two position exchanged two-particle amplitudes in \eqref{eq:ProbabilitySpecialCase} become proportional to simply the amplitude of the unknown particle at different spatial modes $\psi_{u}(y)$ and $\psi_{u}(x)$ and the joint detection probabilities provide information on the symmetric and antisymmetric combinations of these.\\
By expanding \eqref{eq:ProbabilitySpecialCase}
\begin{equation}
%\begin{split}
p_{\alpha,\beta}\left(x,y\right) =  \frac{\left|c\right|^{2}}{4}\left(\left|\psi_{u}(y)\right|^{2}+\left|\psi_{u}(x)\right|^{2}\right) \pm\frac{\left|c\right|^{2}}{2}\left(-1\right)^{\alpha-\beta}\text{Re}\left[\psi_{u}(x)\psi_{u}^{*}(y)\right]\;
%\end{split}
\end{equation}
and writing the position dependent amplitude of the unknown particle as 
\begin{equation}
\psi_{u}(x)=\left|\psi_{u}\left(x\right)\right|\cdot e^{\iE\varphi\left(x\right)}\,,
\end{equation}
with $\varphi\left(x\right)$ representing the local phase profile, we can get access to $\left|\psi_{u}\left(x\right)\right|$ and $\varphi\left(x\right)$ using the relations 
\begin{align}
\left|\psi_{u}\left(x\right)\right| & =  \frac{1}{\left|c\right|}\sqrt{p_{\alpha,1}\left(x,x\right)+p_{\alpha,2}\left(x,x\right)}\nonumber \\
\label{eq:absolute_value}\\
\cos\left[\varphi\left(x\right)-\varphi\left(y\right)\right] & =\pm  \sum_{\beta\in\left\{ 1,2\right\} }\frac{\left(-1\right)^{\alpha-\beta}p_{\alpha,\beta}\left(x,y\right)}{\left|c\right|^{2} \left|\psi_{u}\left(x\right)\right| \left|\psi_{u}\left(y\right)\right|} \label{eq:local_phase}
\end{align}
%\begin{eqnarray}
%\left|\psi_{u}\left(x\right)\right| & = & \frac{\sqrt{2}}{\left|c\right|}\sqrt{p_{\alpha,1}\left(x,x\right)+p_{\alpha,2}\left(x,x\right)}\nonumber \\
%\label{eq:absolute_value}\\
%\cos\left[\varphi\left(x\right)-\varphi\left(y\right)\right] & =\pm & \frac{2}{\left|c\right|^{2} \left|\psi_{u}\left(x\right)\right| \left|\psi_{u}\left(y\right)\right|}\times\nonumber \\
% &  & \sum_{\beta\in\left\{ 1,2\right\} }\left(-1\right)^{\alpha-\beta}p_{\alpha,\beta}\left(x,y\right)\label{eq:local_phase}\\
% &  & \text{for}\,\alpha\in\left\{ 1,2\right\}. \nonumber 
%\end{eqnarray}
for $\alpha\in\left\{ 1,2\right\}$ where by $p_{\alpha,\beta}(x,x)$ we mean $\lim_{y\rightarrow x}p_{\alpha,\beta}(x,y)$. Using these relations, we can learn a lot about the unknown quantum state as represented by $\psi_{u}(x)$. Of course it is not possible to access the global phase of the state (as all states which are identical up to a global phase are physically equivalent). The only missing information is connected to the fact, that $\varphi\left(x\right)-\varphi\left(y\right)$ is not uniquely determined by eq. (\ref{eq:local_phase}). For example\begin{small}
\begin{align}
\varphi\left(x\right)-\varphi\left(y\right) = & \cos^{-1}\left[\pm\sum_{\beta\in\left\{ 1,2\right\} }\frac{\left(-1\right)^{\alpha-\beta}p_{\alpha,\beta}\left(x,y\right)}{\left|c\right|^{2}\left|\psi_{u}\left(x\right)\right|\left|\psi_{u}\left(y\right)\right|}\right]\\
\varphi\left(x\right)-\varphi\left(y\right) = & -\cos^{-1}\left[\pm\sum_{\beta\in\left\{ 1,2\right\} }\frac{\left(-1\right)^{\alpha-\beta}p_{\alpha,\beta}\left(x,y\right)}{\left|c\right|^{2}\left|\psi_{u}\left(x\right)\right|\left|\psi_{u}\left(y\right)\right|}\right]
\end{align}
\end{small}are both valid solutions of eq. (\ref{eq:local_phase}).
As a consequence the states 
\begin{align}
\psi_{u}(x) & = \left|\psi_{u}\left(x\right)\right|\cdot e^{\iE\varphi\left(x\right)}\,,\\
\psi_{u}^{*}(x) & = \left|\psi_{u}\left(x\right)\right|\cdot e^{-\iE\varphi\left(x\right)}
\end{align}
for example are indistinguishable by the interferometric experiment described above.\\
Of course, the above considerations were based on the assumption that we were using a balanced beam splitter and a reference particle of the form $\psi_{r}(x)=c$. However, it is possible to prove, that a similar problem arises for all possible unitary matrices and for all possible reference photons as described by $\vert\psi_{r}\rangle_{2}$.\\
We can show that 
\begin{equation}
U_{\beta1}U_{\alpha2}U_{\beta2}^{*}U_{\alpha1}^{*}\in\mathbb{R}\text{ for }\alpha,\beta\in\left\{ 1,2\right\} \,,\label{eq:real}
\end{equation}
for every unitary matrix $U$. This can be done by the relation
\begin{equation}
\left(\begin{array}{cc}
U_{11}^{*} & U_{21}^{*}\\
U_{12}^{*} & U_{22}^{*}
\end{array}\right)=U^{\dagger}=U^{-1}=\frac{1}{\text{det}\left(U\right)}\left(\begin{array}{cc}
U_{22} & -U_{12}\\
-U_{21} & U_{11}
\end{array}\right).
\end{equation}
Using (\ref{eq:real}) and eq. (\ref{eq:ProbabilityExplicit}), we find that the probability $p_{\alpha,\beta}\left(x,y\right)$ only depends on the absolute value of $\psi_{u}$ and $\psi_{r}$ and the real part of $\psi_{r}(x)\psi_{u}^{*}(x)\psi_{u}(y)\psi_{r}^{*}(y)$ but not on its imaginary part. Hence as long as there is no loss in the system (that is $U$ acting on the port and transverse spatial degrees of freedom is unitary), we can only access $\text{Re}\left[\psi_{r}(x)\psi_{u}^{*}(x)\psi_{u}(y)\psi_{r}^{*}(y)\right]$ and $\left|\psi_{u}(x)\right|^{2}$, which is insufficient to reconstruct the full quantum state. For reconstructing the quantum state $\psi_{u}(x)$, we have to use either at least two different kinds of reference particles or we require additional information on $\psi_{u}(x)$ not accessible by the interferometric setup described above.
\section{State tomography for mixed states\label{sec:State_tomography_mixed_states}}
In the previous section, we have seen that by using only one kind of reference particle $\psi_{r}(x)$, we cannot gain full knowledge of the quantum state $\psi_{u}(x)$. In this section, we show how to overcome this restriction and how to reconstruct the full quantum state of the unknown particle. Here, we go beyond the assumption of the unknown particle being in a pure state and consider mixed quantum states as well. In order to model the corresponding physical situations, we assume that the quantum state of the two particles at the input ports is of the following form
\begin{equation}
\hat{\rho}_{\text{in}}=\hat{\rho}_{u}\otimes\hat{\rho}_{r}\,,
\end{equation}
with the reference particle being again prepared in the pure state
\begin{equation}
\hat{\rho}_{r}=\vert\psi_{r}\rangle_{2}\langle\psi_{r}\vert_{2}\,.
\end{equation}
We can generalise our previous results for pure states to mixed quantum states and obtain the following expression describing the probability of detecting the two particles at the output ports 
\begin{equation}
p_{\alpha,\beta}\left(x,y\right) = {}_{1}\langle\psi_{x,y}^{\alpha,\beta}\vert\hat{\rho}_{u}\vert\psi_{x,y}^{\alpha,\beta}\rangle_{1}\,
\end{equation}
where the detection state is of the same form \eqref{eq:DetectionState1} as in the pure state tomography case.\\
Again we consider a balanced beam splitter (Fig. \ref{fig:Setup}) and a reference particle with a flat profile
\begin{equation}
\psi_{r}(x)=c\,,
\end{equation}
so that the detection state is 
\begin{equation}
\vert\psi_{x,y}^{\alpha,\beta}\rangle_{1} \propto \vert x\rangle\pm(-1)^{\alpha-\beta}\vert y\rangle\,.\label{eq:psi_50_50}
\end{equation}
For the probability densities we obtain
\begin{equation}
%\begin{split}
p_{\alpha,\beta}\left(x,y\right) =  \frac{\left|c\right|^{2}}{4}\left(_{1}\langle x\vert\hat{\rho}_{u}\vert x\rangle_{1}+{}_{1}\langle y\vert\hat{\rho}_{u}\vert y\rangle_{1}\right) \pm\left(-1\right)^{\alpha-\beta}\frac{\left|c\right|^{2}}{2}\text{Re}\left[_{1}\langle x\vert\hat{\rho}_{u}\vert y\rangle_{1}\right]\;.
%\end{split}
\end{equation}
We can use this to extract the real part of the matrix elements of the density matrix 
\begin{equation}
%\begin{split}
\text{Re}\left[_{1}\langle x\vert\hat{\rho}_{u}\vert y\rangle_{1}\right]  = \pm\frac{1}{\left|c\right|^{2}}\left[p_{1,1}\left(x,y\right)-p_{1,2}\left(x,y\right)\right] = \pm\frac{1}{\left|c\right|^{2}}\left[p_{2,2}\left(x,y\right)-p_{1,2}\left(x,y\right)\right]\,.
%\end{split}
\end{equation}
However, $\text{Re}\left[_{1}\langle x\vert\hat{\rho}_{u}\vert y\rangle_{1}\right]$ does not suffice to reconstruct the full quantum state $\hat{\rho}_{u}$, we require information on $\text{Im}\left[_{1}\langle x\vert\hat{\rho}_{u}\vert y\rangle_{1}\right]$ as well. This is connected to the fact that the parallelogram law for a complex Hilbert space is given by
\begin{equation}
\begin{split}
4{}_{1}\langle x\vert\hat{\rho}_{u}\vert y\rangle_{1} =&  \left(\langle x\vert+\langle y\vert\right)\hat{\rho}_{u}\left(\vert x\rangle+\vert y\rangle\right) -\left(\langle x\vert-\langle y\vert\right)\hat{\rho}_{u}\left(\vert x\rangle-\vert y\rangle\right)\\
& -\iE\left(\langle x\vert-\iE\langle y\vert\right)\hat{\rho}_{u}\left(\vert x\rangle+\iE\vert y\rangle\right)\, +\iE\left(\langle x\vert+\iE\langle y\vert\right)\hat{\rho}_{u}\left(\vert x\rangle-\iE\vert y\rangle\right).
\end{split}
\end{equation}
However, this requires not only states $\vert\psi_{x,y}^{\alpha,\beta}\rangle_{1}$ of the form (\ref{eq:psi_50_50}) but also states of the form 
\begin{equation}
\vert\psi_{x,y}^{\alpha,\beta}\rangle_{1}  \propto  \vert x\rangle\pm\iE(-1)^{\alpha-\beta}\vert y\rangle.
\end{equation}
Unfortunately, the states described above correspond to 
\begin{equation}
U_{\beta1}U_{\alpha2}U_{\beta2}^{*}U_{\alpha1}^{*}\in\iE\mathbb{R}\,.\label{eq:condition}
\end{equation}
As outlined in the previous section for a unitary matrix $U$, the above quantity is always real which is in contradiction to the condition (\ref{eq:condition}). Hence, we have to go beyond unitary matrices $U$ in order to get access to $\text{Im}\left[_{1}\langle x\vert\hat{\rho}_{u}\vert y\rangle_{1}\right]\,.$ We can do that by implementing loss or by including an additional degree of freedom, such as polarization.
\section{Photonic implementation of mixed state tomography\label{sec:Photonic_Implementation}}
In the previous section, we found that in order to get full information on the quantum state of the unknown particle, we have to go beyond unitary matrices $U$. In this section, we show how this can be achieved in the case of photons. However, similar considerations also apply to other types of bosonic or fermionic particles. We present two methods by which to obtain a non-unitary matrix $U$, which we describe in the following subsections.
\subsection{Implementation using an additional degree of freedom \label{subsec:Polarization}}
\begin{figure*}
\begin{centering}
\includegraphics[width=6cm]{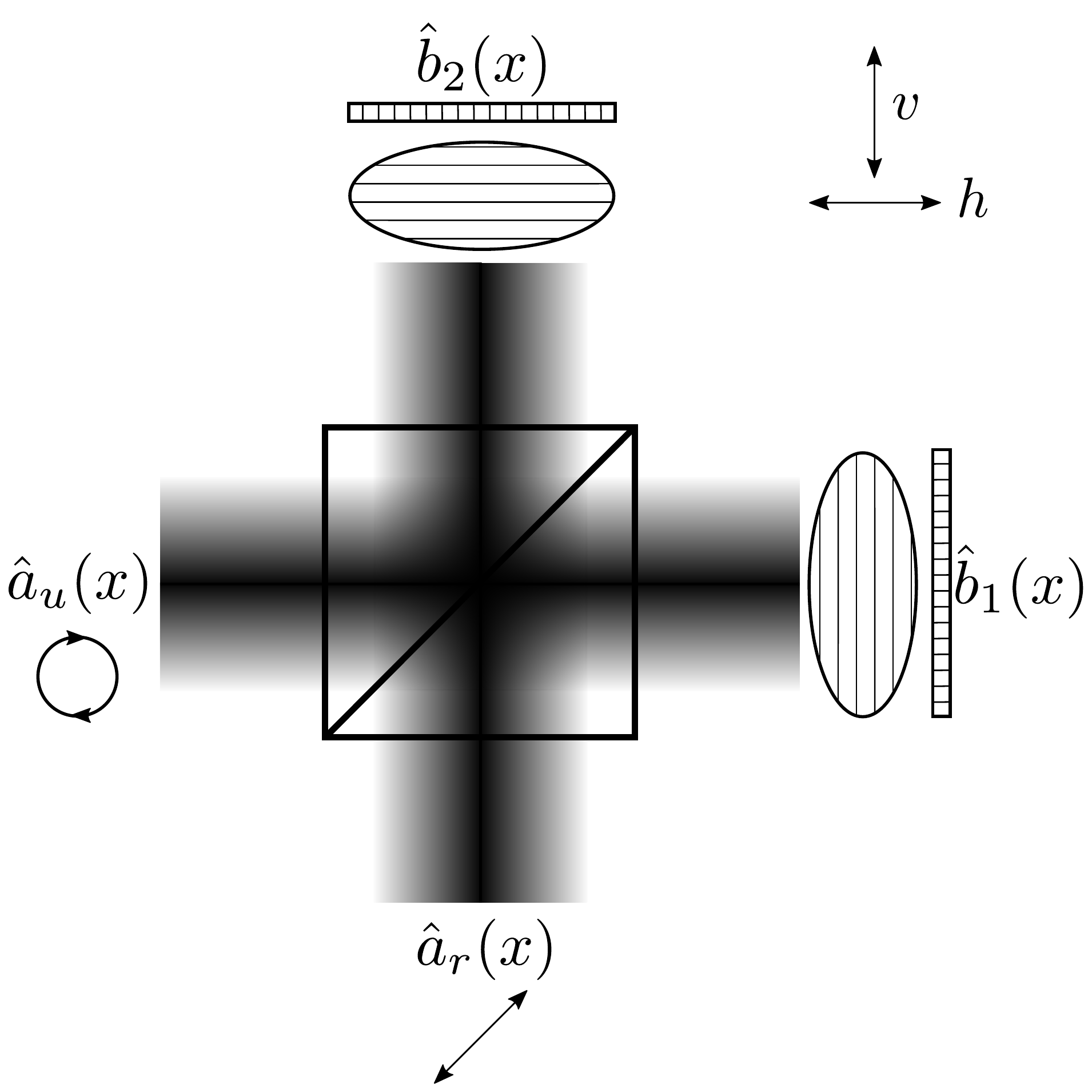} 
\par\end{centering}
\caption{Interferometric setup for performing the mixed state tomography of the photon with the unknown transverse spatial profile by using polarization as additional degree of freedom. In the following we assume that the reference photon at input port $1$ is clockwise circularly polarized and that the unknown photon at input port $2$ is linearly polarized but with a polarization axis which is tilted by $45{\text{\textdegree}}$ with respect to horizontal axis. The polarization filters at the output ports allow us to select and measure observables from a tomographically complete set of observables. \label{fig:Setup-polarization}}
\end{figure*}

In this subsection, we explore the possibility of implementing a non unitary matrix $U$ by including an additional degree of freedom, such as polarization and performing a suitable post-selection. In the following, we assume that the photons at one of the input ports of the beam splitter are prepared in a state of circular polarization and at the other input port the photons are prepared in a state of diagonal polarization while detection at the output ports of the beam splitter happens in the horizontal-vertical basis. The corresponding setup is depicted in Fig. \ref{fig:Setup-polarization}. 
In order to describe the new setup and include the polarization of the photons we have to extend the matrix to a $4\times4$ matrix
\begin{equation}
\left(\begin{array}{c}
\hat{b}_{1h}(x)\\
\hat{b}_{1v}(x)\\
\hat{b}_{2h}(x)\\
\hat{b}_{2v}(x)
\end{array}\right)=\underbrace{\frac{1}{\sqrt{2}}\left(\begin{array}{cccc}
1 & 0 & 1 & 0\\
0 & 1 & 0 & 1\\
1 & 0 & -1 & 0\\
0 & 1 & 0 & -1
\end{array}\right)}_{=U_{\text{BS}}}\left(\begin{array}{c}
\hat{a}_{1h}(x)\\
\hat{a}_{1v}(x)\\
\hat{a}_{2h}(x)\\
\hat{a}_{2v}(x)
\end{array}\right)\,,
\end{equation}
with the indices $h$ and $v$ referring to the horizontal and vertical prioritizations at the input ports. In the following we assume that the unknown photon at input port $1$ is clockwise circularly polarized and that the reference photon at input port $2$ is linearly polarized but with a polarization axis which is tilted by $45^{\text{\textdegree}}$ with respect to horizontal axis. We take this into account, by defining a new set of input modes. The corresponding annihilation operators are connected to the annihilation operators of the horizontally and vertically polarized modes by the relation 
\begin{equation}
\left(\begin{array}{c}
\hat{a}_{1h}(x)\\
\hat{a}_{1v}(x)\\
\hat{a}_{2h}(x)\\
\hat{a}_{2v}(x)
\end{array}\right)  =  \underbrace{\frac{1}{\sqrt{2}}\left(\begin{array}{cccc}
1 & 1 & 0 & 0\\
-\iE & \iE & 0 & 0\\
0 & 0 & 1 & 1\\
0 & 0 & 1 & -1
\end{array}\right)}_{=U_{\text{Basis}}}\left(\begin{array}{c}
\hat{a}_{1\circlearrowright}(x)\\
\hat{a}_{1\circlearrowleft}(x)\\
\hat{a}_{2\nearrow}(x)\\
\hat{a}_{2\searrow}(x)
\end{array}\right)
\end{equation}
Hence we obtain
\begin{equation}
\left(\begin{array}{c}
\hat{b}_{1h}(x)\\
\hat{b}_{1v}(x)\\
\hat{b}_{2h}(x)\\
\hat{b}_{2v}(x)
\end{array}\right) = \underbrace{U_{\text{BS}}\cdot U_{\text{Basis}}}_{=U_{\text{c}}}\left(\begin{array}{c}
\hat{a}_{1\circlearrowright}(x)\\
\hat{a}_{1\circlearrowleft}(x)\\
\hat{a}_{2\nearrow}(x)\\
\hat{a}_{2\searrow}(x)
\end{array}\right)\;,
\end{equation}
with 
\begin{equation}
U_{\text{c}}=\frac{1}{2}\left(\begin{array}{cccc}
1 & 1 & 1 & 1\\
-\iE & \iE & 1 & -1\\
1 & 1 & -1 & -1\\
-\iE & \iE & -1 & 1
\end{array}\right)\;.
\end{equation}
By using our initial condition, we can identify
\begin{equation}
\begin{split}
\hat{a}_{u}(x) & = \hat{a}_{1\circlearrowright}(x)\\
\hat{a}_{r}(x) & = \hat{a}_{2\nearrow}(x)\,.
\end{split}
\end{equation}
Furthermore, we assume that the modes associated to $\hat{a}_{1\circlearrowleft}(x)$ and $\hat{a}_{2\searrow}(x)$ are in the vacuum state. For one set of probability densities we obtain
%\begin{widetext}
\begin{align}
p_{1h,1h}\left(x,y\right) = & \frac{\left|c\right|^{2}}{16}\left(\langle x\vert_{u}\hat{\rho}_{u}|x\rangle_{u}+\langle y\vert_{u}\hat{\rho}_{u}|y\rangle_{u}+2\text{Re}\left[\langle x\vert_{u}\hat{\rho}_{u}|y\rangle_{u}\right]\right)\\
p_{1h,2h}\left(x,y\right) = & \frac{\left|c\right|^{2}}{16}\left(\langle x\vert_{u}\hat{\rho}_{u}|x\rangle_{u}+\langle y\vert_{u}\hat{\rho}_{u}|y\rangle_{u}-2\text{Re}\left[\langle x\vert_{u}\hat{\rho}_{u}|y\rangle_{u}\right]\right)\\
p_{1h,2v}\left(x,y\right) = & \frac{\left|c\right|^{2}}{16}\left(\langle x\vert_{u}\hat{\rho}_{u}|x\rangle_{u}+\langle y\vert_{u}\hat{\rho}_{u}|y\rangle_{u}+2\text{Im}\left[\langle x\vert_{u}\hat{\rho}_{u}|y\rangle_{u}\right]\right)\\
p_{1h,1v}\left(x,y\right) = & \frac{\left|c\right|^{2}}{16}\left(\langle x\vert_{u}\hat{\rho}_{u}|x\rangle_{u}+\langle y\vert_{u}\hat{\rho}_{u}|y\rangle_{u}-2\text{Im}\left[\langle x\vert_{u}\hat{\rho}_{u}|y\rangle_{u}\right]\right).
\end{align}
%\end{widetext}
There are three more such sets of probability densities. These can be obtained by \textsc{i}) replacing $h$ with $v$ and vice versa \textsc{ii}) replacing 1 with 2 and vice versa and \textsc{iii}) joint application of \textsc{i}) and \textsc{ii}). These however contain exactly the same tomographic information as the set written out explicitly above. If the polarization measurements at each output of the beam splitter are done with a polarizing beam splitter with a detector at each of its outputs then one has access to all four of the above output probability densities. In this case one can combine them rather straightforwardly to obtain an expression for the matrix element
\begin{equation}
%\begin{split}
\frac{\vert c \vert^2}{4} \langle x\vert_{u}\hat{\rho}_{u}|y\rangle_{u} = p_{1h,1h}(x,y) - p_{1h,2h}(x,y)-\iE p_{1h,1v}(x,y) +\iE p_{1h,2v}(x,y).
%\end{split}
\end{equation}
If, however,  for some reason one can only do the polarization measurements with a polarizer and a detector at each of the outputs of the beam splitter then access to the last probability density is not possible. Even in this case complete tomography is possible as the first three relations already provide enough information for the reconstruction of the density matrix element
\begin{equation}
%\begin{split}
\frac{\vert c\vert^2}{4}\langle x\vert_{u}\hat{\rho}_{u}|y\rangle_{u} = (1-\iE)p_{1h,1h}(x,y) - (1+\iE)p_{1h,2h}(x,y)  + 2\iE p_{1h,2v}(x,y).
%\end{split}
\end{equation}

\subsection{Implementation using loss \label{subsec:Loss}}
\begin{figure*}
\begin{centering}
\includegraphics[width=7cm]{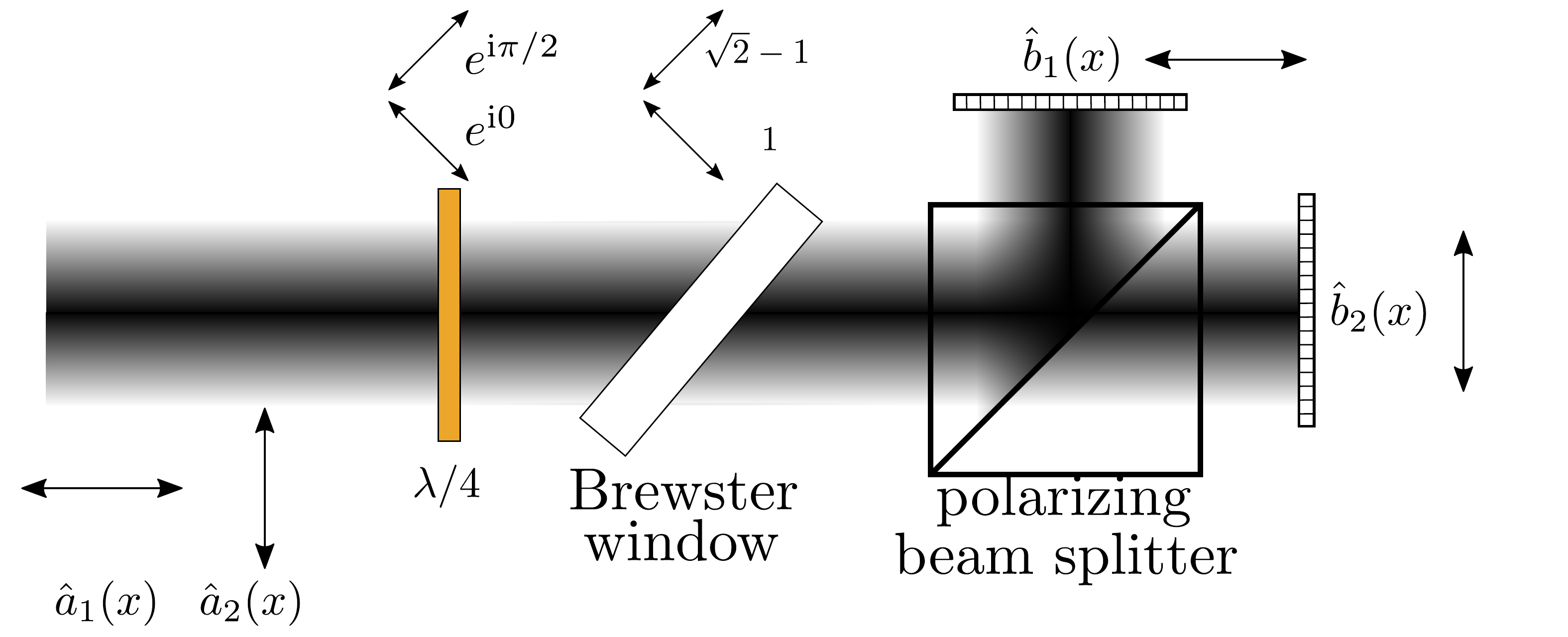} 
\par\end{centering}
\caption{Interferometric setup for performing the mixed state tomography of the photon with the unknown transverse spatial profile by carefully engineering loss in the system. Note that here both the unknown and reference photons enter in the same port but different polarization modes.
The loss in the system can be tuned by varying the orientation of the Brewster window.
The Brewster window allows us to choose different loss rates for two orthogonal polarizations. \label{fig:Loss}}
\end{figure*}

The second possibility to reconstruct the quantum state, is to introduce losses to the system, which are described by a suitable non unitary matrix $U$. This possibility is explored in this subsection. In order to obtain a suitable matrix $U$ the losses introduced to the system have to be engineered carefully. A possible implementation is depicted in Fig. \ref{fig:Loss}. In the following, we choose the basis element $\mathbf{e}_{1}$ to represent the horizontal polarization and $\mathbf{e}_{2}$ for the vertical polarization. A rotation is described by the following matrix 
\begin{equation}
U_{\text{rot}}(\theta)=\left(\begin{array}{cc}
\cos\theta & \sin\theta\\
-\sin\theta & \cos\theta
\end{array}\right)
\end{equation}
the $\lambda/4$-plate is modelled by the matrix 
\begin{equation}
U_{\lambda/4}=\left(\begin{array}{cc}
1 & 0\\
0 & \iE
\end{array}\right)
\end{equation}
and the Brewster window \cite{schleich2011quantum} is represented by the non-unitary matrix 
\begin{equation}
U_{\text{B}}\left(\eta\right)=\left(\begin{array}{cc}
1 & 0\\
0 & \eta
\end{array}\right)
\end{equation}
with $\eta$ being the damping factor of the vertical polarized component of the beam. In order to get full information on the quantum state we will see, that $0<\eta<1$. The damping factor $\eta$ can be tuned by varying the angle of the Brewster window. The setup in Fig. \ref{fig:Loss} is described by 
\begin{equation}
U_{\text{B}}\left(\eta\right)U_{\lambda/4}=\left(\begin{array}{cc}
1 & 0\\
0 & \iE\eta
\end{array}\right)\,.
\end{equation}
By choosing $\hat{a}_{1}(x)$ and $\hat{a}_{2}(x)$ to represent the annihilation operators of the input modes with respect to the polarizations as depicted in Fig. \ref{fig:Loss} and $\hat{b}_{1}(x)$ and $\hat{b}_{2}(x)$ to be the output ports of the beam splitter, we obtain 
\begin{equation}
\left(\begin{array}{c}
\hat{b}_{1}(x)\\
\hat{b}_{2}(x)
\end{array}\right)=U_{\text{rot}}\left(\pi/4\right)U_{\text{B}}\left(\eta\right)U_{\lambda/4}U_{\text{rot}}\left(-\pi/4\right)\left(\begin{array}{c}
\hat{a}_{1}(x)\\
\hat{a}_{2}(x)
\end{array}\right)\,.
\end{equation}
For $\eta=\sqrt{2}-1$ we obtain
%\begin{widetext}
\begin{small} 
\begin{equation}
U=U_{\text{rot}}\left(\pi/4\right)U_{\text{B}}\left(\eta\right)U_{\lambda/4}U_{\text{rot}}\left(-\pi/4\right)=\frac{1}{2}\iE\left(\sqrt{2}+(-1-\iE)\right)\left(\begin{array}{cc}
1 & \left(\iE-1\right)/\sqrt{2}\\
\left(\iE-1\right)/\sqrt{2} & 1
\end{array}\right)\,.
\end{equation}
\end{small}
%\end{widetext}
As it turns out, the above matrix satisfies condition (\ref{eq:condition}) and allows us to get access to $\text{Im}\left[_{1}\langle x\vert\hat{\rho}_{u}\vert y\rangle_{1}\right]$.
By rotating the $\lambda/4$-plate or by changing $\eta$ of the Brewster window, we can measure a complete set of observables for performing the state tomography. For example, by removing the Brewster window or choosing an angle such that $\eta=1$ we recover the lossless case obtaining
\begin{equation}
U=U_{\text{rot}}\left(\pi/4\right)U_{\text{B}}\left(1\right)U_{\lambda/4}U_{\text{rot}}\left(-\pi/4\right)=\frac{1+\iE}{2}\left(\begin{array}{cc}
1 & \iE\\
\iE & 1
\end{array}\right)\,.
\end{equation}
This $U$ indeed satisfies \eqref{eq:real} and thus allows us to get access to $\text{Re}\left[_{1}\langle x\vert\hat{\rho}_{u}\vert y\rangle_{1}\right]$. By combining the information on $\text{Re}\left[_{1}\langle x\vert\hat{\rho}_{u}\vert y\rangle_{1}\right]$ and $\text{Im}\left[_{1}\langle x\vert\hat{\rho}_{u}\vert y\rangle_{1}\right]$, we can reconstruct the full density matrix $\hat{\rho}_{u}$.
\section{Conclusion\label{sec:Conclusion}}
In summary, we have analysed the application of two particle interference as a tool for single particle tomography. We focused in particular on the transverse spatial profile of single particles. By using two particle interference, we have characterized the amplitude as well as the local phase variations of single particle states. Retrieving local phase variations of single particle states constitutes a challenging task, as the global phase of single particle states is undefined. Tomographic methods based on two particle interference, help to overcome this obstacle and allow us to extract full information about the local phase variations.\\
Our theoretical discussion is based on the framework of second quantization. This enabled us to develop protocols for the tomography of bosonic or fermionic particles. Hence, our protocols can be used to perform single particle tomography on bosonic particles such as photons and bosonic atoms as well as fermionic particles such as neutrons and fermionic atoms. This allowed us to go beyond the imaging technique applied in \cite{HOMTomography} for reconstructing the phase profile of a pure single photon state. Furthermore, we have generalized the method to the tomography of mixed states. We have shown that by carefully engineering loss or taking additional degrees of freedom into account our method can be used to gain access to observables whose measurements provide tomographically complete information about mixed single particle quantum states.\\
In addition to our theoretical discussion, we have developed two proposals of interferometric setups for performing the tomography of the transverse spatial profile of mixed single photon states. However, similar considerations can also be applied to other types of indistinguishable particles such as bosonic or fermionic atoms or neutrons.
\section*{Acknowledgements}
We thank Gernot Alber for helpful comments and suggestions. This work was supported by the DFG as part of the \textsc{CRC} 1119 \textsc{CROSSING}, the EPSRC through the quantum hub {Q}uant{IC} \textsc{EP/M01326X/1} and the Royal Society.
%\end{multicols}
%{unsrt} %apsrev4-1
%\bibliographystyle{apsrev4-1}
%\bibliography{RetroBib}
\printbibliography

\end{document}